\documentclass[showkeys,showpacs,amsmath,superscriptaddress,preprintnumbers,amssymb,10pt,twocolumn,pra]{revtex4-2}
\usepackage[T1]{fontenc}
\usepackage{amsmath}
\usepackage{graphicx}
\usepackage{graphics}
\usepackage{graphpap}
\usepackage{epstopdf}
\usepackage[dvipsnames]{xcolor}
\usepackage{float}
\usepackage{amssymb}
\usepackage{bm}
\usepackage{color}
\usepackage{xcolor}
\usepackage{subfigure}
\usepackage{ulem}
\usepackage{soul}
\usepackage{braket}

\begin{document}
	\title{Simulation of single-qubit gates in spin-orbit coupled Bose-Einstein condensate with  cubic-quintic nonlinearity by nonlinear perturbations}
	\author{Prithwish Ghosh}
	\email{prithwishghose@gmail.com}
	\affiliation{Department of Physics, Kazi Nazrul University, Asansol-713340, W.B., India}
	\author{ Kajal Krishna Dey}
	\email{kajaldeypkc@gmail.com}
	\affiliation{Department of Physics, Banwarilal Bhalotia College, Asansol-713303, W.B., India}
	\author{ Golam Ali Sekh}
    \email{skgolamali@gmail.com}
	\affiliation{Department of Physics, Kazi Nazrul University, Asansol-713340, W.B., India}

\begin{abstract}
We consider spin-orbit coupled Bose–Einstein condensates with cubic-quintic nonlinear interaction within the framework of second quantization formulation and find eigen states using numerical simulation and mean-field approximation. We show that two low-lying  Schr\"odinger cat states remain degenerate upto a certain value of Raman coupling strength and these states can  serve as qubit basis. We take three different nonlinear perturbations and find that the perturbations can result in different rotations of qubit state on Bloch sphere. We calculate the unitary operator corresponding to each perturbation and suggest the possibilities for obtaining various gates in ultra-cold atomic system.	\end{abstract}
\maketitle

\section{Introduction}
When a dilute gas of bosons is cooled  to near absolute zero temperature, it causes the bosons to occupy the lowest energy level and creates a macroscopic quantum state known as Bose-Einstein condensate (BEC) \cite{Klaers, Bunkov}. Due to the collective nature and coherence property  of a BEC, it is found suitable for quantum computation, quantum metrology and quantum simulation. In comparison with single-particle systems, qubits encoded in BECs either by occupying different condensate modes or by internal hyperfine states are benefited from the variable interaction strengths and enhanced robustness against particle loss \cite{Chen, Tian, Kapale, Makhlin, Adrianov, Mohseni, Ghosh}.

Early proposals have demonstrated the feasibility of encoding quantum information in two-mode BECs using hyperfine states, laying the foundation for gate operations through controlled evolution of the condensate wave function \cite{Byrnes}. Neutral atoms confined in optical lattices or microtraps are another well-studied platform for quantum computing. In this case, qubits are determined by either spatial localization or internal atomic states and, spin-dependent potentials or collisional dynamics are used to realize gate operations \cite{Schneider, Elisha}. Ensemble-based qubits, which are made up of numerous atoms in a common quantum state, provide improved stability against particle loss and lower Rabi frequency requirements, whereas single-atom qubits provide precise addressability \cite{Brion2, Beterov, Lukin}. Atomic BEC offers a potential platform for achieving entangled qubit ensembles and quantum correlations \cite{Klaers, Bunkov}. Because of their enormous occupation numbers and coherent phase features, BEC-based qubits can make read-out procedures simpler. 

Nonlinear quantum gates, which are known to have computational advantages over linear gate sets in specific contexts \cite{Adrianov}, can be implemented by utilizing the intrinsic nonlinearity of BECs due to atomic interaction. Higher-order interaction systems, such as cubic-quintic nonlinearities, provide new avenues for designing efficient qubit landscapes in which particular eigenstates can be isolated or degenerated according to interaction parameters. {Higher-order interaction strengths can be tuned by applying external magnetic field using  Feshbach resonance \cite{Yuce} as the scattering length is tunable with external magnetic field.} The coherent control of single-qubit BECs has already been demonstrated on atom chips \cite{Böhi, Riedel}, and recent advances have shown entanglement between spatially separated BECs, a key step toward implementing multi-qubit architectures \cite{Julsgaard}. Additionally, these devices provide pathways for quantum teleportation methods that make use of macroscopic spin ensembles \cite{Krauter, Pyrkov}.

Our objective in this work is to investigate how two lowest energy states of spin-orbit coupled  Bose-Einstein condensate with cubic-quintic nonlinear interactions can be used to encode qubits. The energies of low-lying eigen states coupled by spin-orbit interaction vary with Raman coupling and atomic interaction parameters. For given values of atomic interaction, the lowest energy of the system decreases with Raman coupling. However, two lowest energy states appear to be nearly degenerate for a certain range of Raman coupling parameter. These two low-lying states can be used to generate qubit states. More specifically, the Schr\"odinger cat states obtained through mean-field approximation provide the basis of qubit state.  One can achieve dynamic control of qubit features through tuning the strength of spin-orbit coupling  and   nonlinearity.  The degenerate ground states occurred due to interplay between spin-orbit coupling  and quintic interaction may provide a possible path toward all-atomic quantum logic devices.

The structure of the paper is organized as follows. In section II, we first analyze the many-body Hamiltonian using exact diagonalization to find a quasi-degenerate low-energy subspace. We then develop a mean-field description to obtain analytical insight into the origin of this degeneracy and derive the relevant control parameter. Next, we use beyond mean-field theory to construct symmetric and antisymmetric superpositions of the mean-field states, which define the effective qubit basis. In section III, we introduce controlled perturbations and demonstrate how they can be used to implement single-qubit gate operations. Finally, in section IV,  we give concluding remarks.

\section{Theoretical formulation}
We consider a two-level Bose-Einstein condensate coupled through linear and nonlinear interactions. The nonlinear interaction can arise from  interactions between the atoms in the same (intra-atomic) as well as different (inter-atomic) levels and the linear interaction arise from spin-orbit coupling.{This type of coupled system can be described by the following three-dimensional Gross-Pitaevski(GP) equation \cite{r22}
\begin{equation}
i \hbar \frac{\partial \hat{\Psi}(\vec{r},t)}{\partial t} = \bigg[\frac{1}{2m} (\hat{p} + \hat{A})^2+\frac{\delta}{2} \boldsymbol{\sigma_z}+ \hat{V}(r) + \hat{H}_{int}\bigg] \hat{\Psi}(\vec{r},t).
\end{equation}
Here $\hat{\Psi}(x,t)$ is the field operator, $\delta$ is the tunable Zeeman detuning parameter and $\hat{H}_{int}$ stands for interaction Hamiltonian matrix. The external potential, $\hat{V}(r)=\hat{v}(x)+\frac{m}{2}\big[\omega_\perp^2(y^2+z^2)\big]\boldsymbol{I}$ with $\hat{v}(x)= \frac{m}{2}\omega_x^2 x^2 \boldsymbol{I}$,  can be experimentally realized either by a magnetic trap or an optical trap \cite{ Pritchard, Hess}}. The spin-orbit coupling (SOC) is created through the generation of non-abelian gauge potential and the coupling between the condensates in the two levels is controlled using appropriate Raman laser beams. For equal weights of Rashba and Dresselhaus SOCs, the gauge potential is given by \cite{Zhai}
\begin{equation}
	\hat{A}= k_0 \boldsymbol{\sigma_x},
\end{equation}
where $k_0={2 \pi}/{\lambda}$ is the wave vector of the laser and $\boldsymbol{\sigma_x}$ is Pauli spin matrix.

{Considering the frequency of transverse trap ($\omega_\perp $) much larger than that ($\omega_x$) of longitudinal trap and thus  freezing the transverse motion, the ground state can be taken as $\hat{\Psi}(\vec{r},t)=\hat{\psi}(x,t)\phi_0(y,z)$, where $\phi_0(y,z)=\big(\frac{m \omega_\perp}{\pi \hbar}\big)^{1/2} \exp{[-m\omega_\perp(y^2+z^2)}/(2 \hbar)]$.
Multiplying both sides of Eq. (1) by $\phi_0^*$ and then integrating over the transverse variables $y,\, z$ we obtain the following quasi-one-dimensional field equation}
\begin{equation}
i \hbar \frac{\partial \hat{\psi}(x,t)}{\partial t} = \bigg[\frac{1}{2m} (\hat{p}_x +  \hat{A})^2+\frac{\delta}{2} \sigma_z+ \hat{v}(x) + \hat{H}_{int}\bigg] \hat{\psi}(x,t).
\end{equation}
Here $\hat{\psi} = [\hat{\psi}_{\uparrow},\hat{\psi}_{\downarrow}]^{T}$ stands for the two-component field operator. Particularly, the order parameters of the condensates associated with  two levels are denoted by $\hat{\psi}_{\uparrow}$ and $\hat{\psi}_{\downarrow}$ respectively and they are connected to annihilation operators $\hat{a}$ and $\hat{b}$ through $\hat{\psi}_{\uparrow}=\psi_0(x) \hat{a} $ and $ \hat{\psi}_{\downarrow}=\psi_0(x) \hat{b}$ with $\psi_0(x)$, the ground state wave function. We re-scale $t$ by $t \omega_\perp$ and $x$ by $x a_\perp^{-1}$ in Eq.(3) and write interacting Hamiltonian  as follows \cite{Zhang},
\begin{equation}
		\hat{H}_{int}=\begin{bmatrix}
			\hat{h}_1 & 0\\
			0 & \hat{h}_2.
		\end{bmatrix}
\end{equation}
Here {$\hat{h}_1=(g_{{\uparrow}{\uparrow}} |\hat{\psi}_{\uparrow}|^2 + g_{{\uparrow}{\downarrow}} |\hat{\psi}_{\downarrow}|^2 + h_{{\uparrow}{\uparrow}}|\hat{\psi}_{\uparrow}|^4)$, 
$\hat{h}_2=(g_{{\uparrow}{\downarrow}} |\hat{\psi}_{\uparrow}|^2 + g_{{\downarrow}{\downarrow}} |\hat{\psi}_{\downarrow}|^2 + h_{{\downarrow}{\downarrow}}|\hat{\psi}_{\downarrow}|^4)$ with $h_{{\downarrow}{\downarrow}}=h_{{\uparrow}{\uparrow}}$ and $g_{{\downarrow}{\downarrow}}=g_{{\uparrow}{\uparrow}}$ give  intra-atomic quintic and cubic interactions respectively  while  $g_{{\uparrow}{\downarrow}}$ gives inter-atomic cubic interaction. {They are connected with interaction parameters ($g_1, g_2 \,{\rm and }\,g_{12}$) in three-dimension as
$ g_{\uparrow \uparrow }=\frac{g_1}{2 \pi \hbar \omega_\perp a_\perp^2} , h_{\uparrow \uparrow}=\frac{g_2}{3 \pi^2 \hbar \omega_\perp a_\perp^4}$ and  $g_{\uparrow \downarrow }=\frac{g_{12}}{2 \pi \hbar \omega_\perp a_\perp^2}$ where $a_\perp=\sqrt{\frac{\hbar}{m \omega_\perp}}$.}

In the second quantization formulation, total Hamiltonian of system is given by
\begin{equation}
		\hat{H}_0= \langle \psi|\hat{H}|\psi\rangle.
\end{equation}
To understand the ground state property of the system, we describe Eq. (5) by the following two-mode model 
\begin{eqnarray}
\hat{H}_0&=&u_0(\hat{a}^{\dagger}\hat{a}^{\dagger}\hat{a}^{\dagger}\hat{a} \hat{a}\hat{a} + \hat{b}^{\dagger}\hat{b}^{\dagger}\hat{b}^{\dagger}\hat{b} \hat{b} \hat{b}) +2 u_2 \hat{a}^{\dagger} \hat{b}^{\dagger} \hat{b} \hat{a} \nonumber\\&+&  (u_0 + u_1) (\hat{a}^{\dagger} \hat{a}^{\dagger} \hat{a} \hat{a} +\hat{b}^{\dagger} \hat{b}^{\dagger} \hat{b} \hat{b}) +  (\hat{a}^{\dagger} \hat{a} - \hat{b}^{\dagger} \hat{b}) \frac{\delta}{2} \nonumber\\&+&  {k_0 {p}_0}( \hat{a}^{\dagger} \hat{b} +  \hat{b}^{\dagger} \hat{a}).
\end{eqnarray}
Here $u_0=\int_{}^{}h_{{\uparrow}{\uparrow}}|\psi_0|^6 dx$ represents the quintic interaction due to the attractive three-body interactions in the two-component BEC \cite{Hammond}, $u_1=\int_{}^{} g_{{\uparrow}{\uparrow}} |\psi_0|^4 dx$ and $u_2=\int_{}^{} g_{{\uparrow}{\downarrow}} |\psi_0|^4 dx$ stand for the effective values of intra- and inter-atomic two-body interactions, and $p_0=\langle \psi_0|\hat{p}_x |\psi_0\rangle/m $. Understandably, the creation operators, $a^{\dagger}$ and $b^{\dagger}$ create particles in $|\uparrow\rangle$ and $|\downarrow\rangle$ respectively. In Eq.(6), $\hat{H}_0$ represents a Hamiltonian in Fock space and the states are linked through the term $k_0 {p}_0(\hat{a}^{\dagger}\hat{b} + \hat{b}^{\dagger}\hat{a})$. It matches with the standard two-mode expansion of a two-component BEC in two dressed spin states.

{Note that Eq.(3) describes two-component Bose-Einstein condensates with cubic -quintic nonlinear interactions. This interaction can be  controlled using Feshbach resonance technique \cite{Volz, Inouye, Fatemi}. The quintic interaction arises due to three-body interaction and it diminishes  if the distance of third particle is far from other two particles.  Particularly, quintic nonlinearity becomes important for larger value of s-wave scattering length \cite{Zhang,Altin,Hamid}.  In a typical  $^{87}$Rb experiment with  $g_1 \approx 5 \hbar \times 10^{-17} m^3 s^{-1}$, $g_{12}\approx 5.26 \hbar \times 10^{-17} m^3 s^{-1}$  and $g_2 \approx 4\hbar \times 10^{-38} m^6 s^{-1}$ \cite{Zhang,Altin}, the quintic and cubic interaction-parameters can be calculated for $\omega_\perp\sim 2\pi\times 100$ rad\,s$^{-1}$  and $\omega_x\sim 2\pi\times 4k$ rad\,s$^{-1}$ as $u_0 \approx 3.35 \times 10^{-6}$, $u_1 \approx 2.85 \times 10^{-3}$ and $u_2 \approx 3.0 \times 10^{-3}$.}

For $N$ number of particles, the Hamiltonian in Eq.(6) is a $(N+1)\times (N+1)$ dimensional matrix and it can be solved numerically taking moderate value of $N$. We diagonalize  the matrix and  calculate energies of ground state ($E_0$) and excited states ($E_i$) considering negligible Zeeman detuning. It is seen that $E_0$ is larger for smaller values of $k_0$ (Fig.1). However, the differences of energy between the excited and the ground states show some interesting features. Particularly, the variation of the energy difference of different excited states from the ground state clearly  shows that,  below a critical value of $k_0/N$,  two lowest energy states are quasi-degenerate (Fig.2). In addition, these two states are well separated from the rest of the spectrum. We have checked that this degeneracy is maintained for a very weak Zeeman splitting and the degeneracy breaks if  Zeeman splitting increases. The quasi-degeneracy between the two lowest many-body eigenstates arises from the competition among three energy scales. They are Raman (spin-flip) coupling, two-body (cubic) interactions and three-body (quintic) interactions. { Within the two-mode approximation, the effective Hamiltonian contains terms scaling differently with atom number $N$ as, $E_{\rm cubic} \sim u_1 N(N-1)$, and $E_{\rm quintic} \sim u_0 N(N-1)(N-2)$. Thus, higher-order interactions acquire enhanced weight in mesoscopic condensates due to their stronger $N$-scaling. This indicates that attractive quintic interaction can stabilize the two-minima structure required for Schr\"{o}dinger-cat state formation.}

\begin{figure}[h!]
\centering
\includegraphics[width=8.0cm, height=5.5cm]{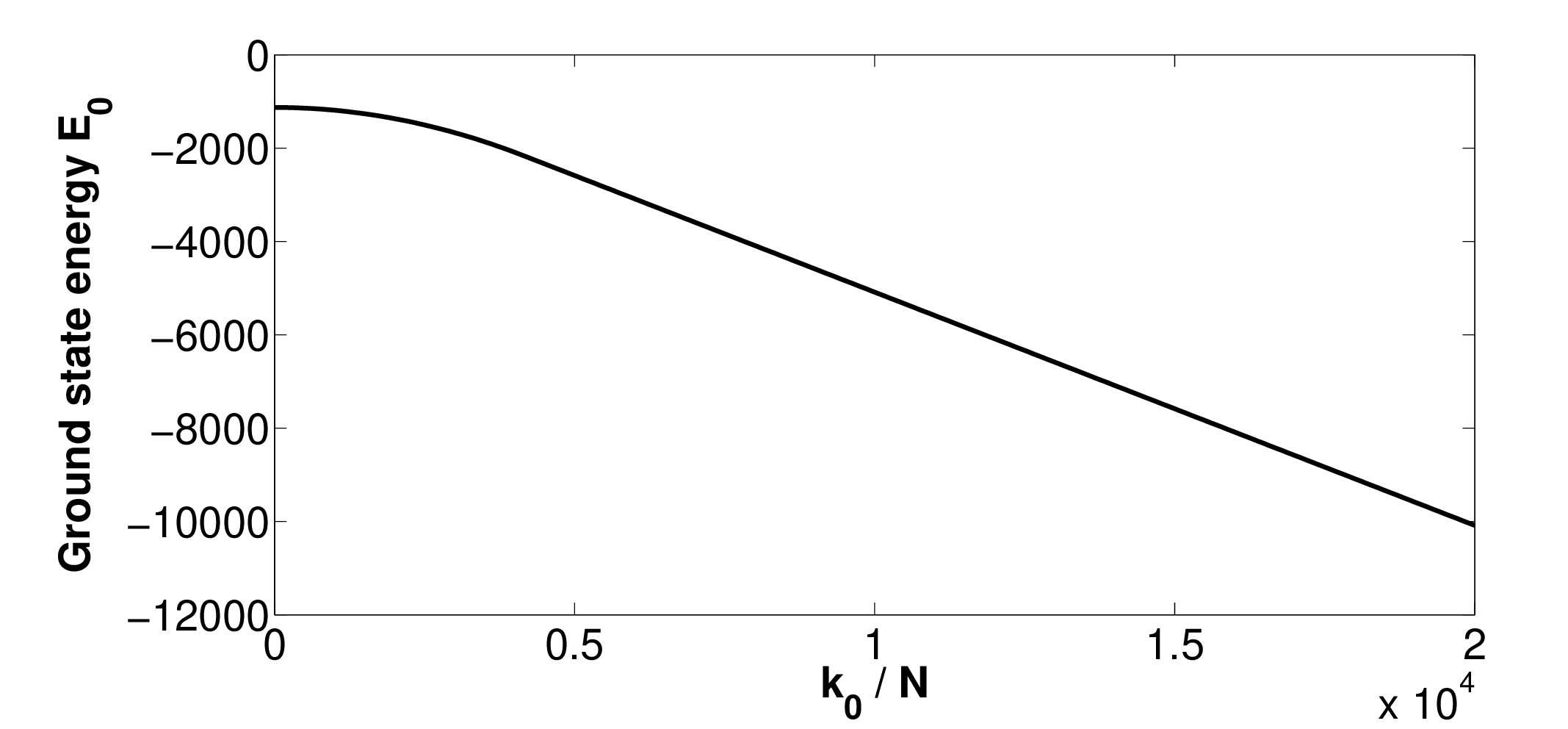}
\caption{The ground state energy as a function of $k_0/N$ for the set of parameter values $u_0 = -3.35\times 10^{-6}, u_1= -2.85\times 10^{-3}, u_2 = 3.0\times 10^{-3}, N=500$ and ${p}_0= -1\times 10^{-6}$ obtained from exact calculation.}
\label{fig1}
\end{figure}
\begin{figure}[h!]
\centering
\includegraphics[width=8.0cm, height=5.5cm]{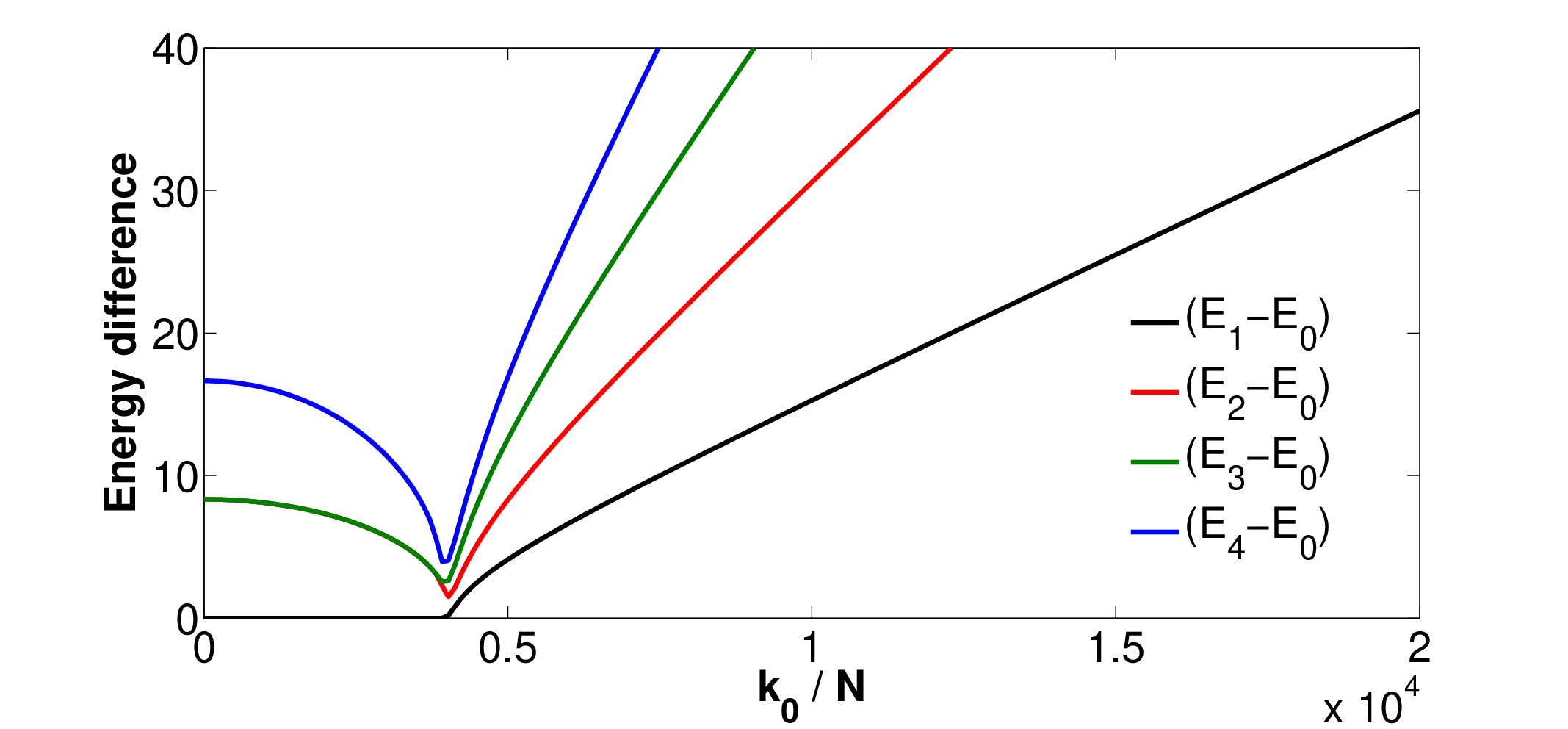}
\caption{The energy difference of low-lying excited states from the ground state as a function of $k_0/N$ for the set of parameter values $u_0 = -3.35\times 10^{-6}, u_1= -2.85\times 10^{-3}, u_2 = 3.0\times 10^{-3}, \delta=0, N=500$ and $p_{0}= -1\times 10^{-6}$ obtained from exact calculation.}
\label{fig2}
\end{figure}
{We see that the exact diagonalization of the Hamiltonian displays the quasi-degenerate low-lying states, but the dependence of the degeneracy on system parameters such as Raman coupling strength and interaction coefficients remains difficult to interpret directly from numerical results. To get the analytical insight into the two-component SOC BEC, we now turn to a mean-field description based on a spin-coherent state ansatz.}

\section{Mean field approximation for low-lying eigen state}
 
Spin-orbit coupled Bose-Einstein condensate is a two-level system where atoms are occupying two internal states (pseudo-spin states). The most general class of single particle states  of the system is constructed through arbitrary superposition  of internal states, namely,  $|\uparrow \rangle$ and $|\downarrow\rangle$. In the mean-field formalism, all the atoms occupy the same state. The collective state of $N$ particles is described by the  following ansatz  \cite{Byren}
\begin{eqnarray}
|\psi_N\rangle = \frac{1}{\sqrt{N!}}(\alpha \hat{a}^{\dagger} + \beta \hat{b}^{\dagger})^N |vac \rangle.
\end{eqnarray}
This class of state is called a spin-coherent state and plays a central role in manipulations of BECs. From the normalization condition, Eq. (7) gives $|\alpha|^2+|\beta|^2=1$. The energy corresponding to this ansatz obtained from $\langle \psi_N |\hat{H}_0|\psi_N \rangle$ is given by 
\begin{eqnarray}
E(\alpha,\beta)&=&u_0N(N-1)(N-2)(|\alpha|^6+|\beta|^6) \nonumber\\&+&(u_0+u_1)N(N-1)(|\alpha|^4 +|\beta|^4)\nonumber\\&+&2u_2N(N-1)|\alpha|^2|\beta|^2  + N k_0 p_0  (\alpha^* \beta\nonumber\\&+& \alpha \beta^*) + \frac{\delta}{2} N(|\alpha|^2-|\beta|^2).
\end{eqnarray}
Here $N$ is the number of particles for the system. Considering $\alpha$ and $\beta$ as  real parameters, we optimize the energy  $E(\alpha,\beta)$ using Lagrange undetermined multiplier  for zero Zeeman splitting and  get following roots.
\begin{equation}
\alpha_{\pm} = \beta_{\pm} = \bigg[ \frac{1}{2}(1 \pm \sqrt{1-\Lambda^2})\bigg]^{1/2}
\end{equation}
with
\begin{equation}
\Lambda=\frac{2 k_0 p_0}{(N-1)[3u_0(N-2)+2(u_0+u_1-u_2)]}.
\end{equation}
Understandably, $\Lambda \le 1$ for real values of $\alpha$ and $\beta$.
Quasi-degeneracy of the two  states occurs in the regime $\Lambda \le 1$, where two macroscopically distinct occupation configurations coexist. { To quantify the relative role of cubic and quintic nonlinearities, we consider the ratio
\begin{eqnarray}
	R = \frac{3u_0 (N-2)}{2(u_0 + u_1 - u_2)} \nonumber
\end{eqnarray} 
For mesoscopic condensates $(N \gg 1)$,
\begin{eqnarray}
	R \approx \frac{3 u_0 N}{2(u_1 - u_2)} \nonumber
\end{eqnarray} 
Even when $|u_0|\ll|u_1|$, the multiplicative $N$-enhancement can make the quintic contribution comparable to or larger than the cubic one.}
\begin{figure}[h!]
	\centering
	\includegraphics[width=8.0cm, height=5.0cm] {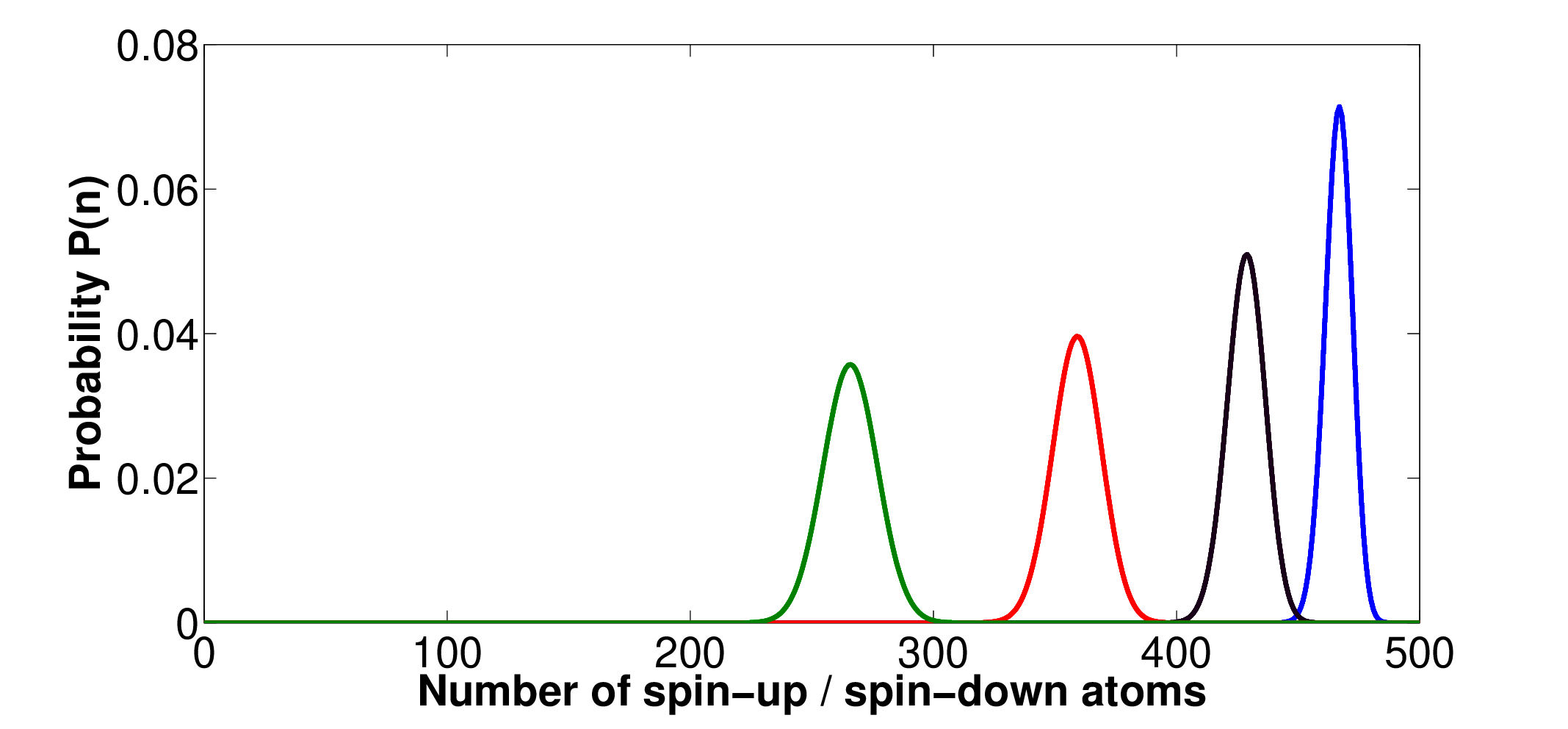}
	\caption{Probability distribution for the ground state as derived from mean-field theory as a function of spin-up atom number. $\Lambda=0.5$ (blue), $\Lambda=0.7$ (black), $\Lambda=0.9$ (red) and $\Lambda=0.998$ (green) for $N=500$.}
\label{fig3}
\end{figure}
In order to understand the role of $\Lambda$ on occupation of the two-level system $(\Ket\uparrow,\Ket\downarrow)$ for $\{\alpha=\alpha_{+},\beta=\beta_{+}\}$, we rewrite $\Ket{\psi_N}=\frac{1}{\sqrt{N!}}\sum_{N_\uparrow=0}^N\sqrt{^NC_{N_\uparrow}}\alpha^{N_\uparrow}\beta^{N_\downarrow}|N_\uparrow,N_\downarrow\rangle$ and express the probability of occupation of  $|\uparrow\rangle$ in terms of $\Lambda$ as $P_\uparrow=|\sqrt{^NC_{N_\uparrow}}\alpha^{N_\uparrow}(\Lambda)\beta^{N_\downarrow}(\Lambda)|^2$, where $N_\uparrow+N_\downarrow=N$.  In Fig.3 we display the analytic results of  occupation probability ($P$) as a function of $N_\uparrow$ taking $N=500$. 

There is no laser coupling between two internal states of the atoms when $\Lambda=0$. As a result, only the collisional interactions determine the structure of the ground state, and the last two terms of the Hamiltonian disappear. In this regime, the interaction energy is minimized when all atoms are either in the spin-up state or in the spin-down state. In the intermediate regime $0 < \Lambda < 1$, the interaction terms of $\hat{H}_0$ compete with the Raman coupling. For $\Lambda= 1$, the Raman coupling term is the dominating term and the energy of the coupling term is minimized if each atom is in an equal superposition between spin-up and spin-down states. The mean field ground state fluctuates constantly between the two extreme cases for $\Lambda = 0$ and $\Lambda =1$. {If we define population imbalance as $z=|\alpha|^2-|\beta|^2$, then for large $N$ values the mean field energy per particle reduces to $E(z)\approx U_{\rm eff} z^2-2 k_0 p_0 \sqrt{(1-z^2)}$, where $U_{\rm eff}=3u_0 (N-2)/2 + (u_0 + u_1 -u_2)$. In the regime $|\Lambda| \le 1$, the energy develops two degenerate minima at $z=\pm\sqrt{1-\Lambda^2}$ corresponding to two macroscopically distinct states. However, the mean-field ansatz selects only one of these minima, thereby breaking the underlying symmetry of the Hamiltonian. The exact many-body ground state must reveal this symmetry and is therefore expected to be a coherent superposition of the two mean-field solutions. So, it is necessary to go beyond mean-field theory and construct symmetric and antisymmetric superpositions of the two macroscopically occupied states.}

\section{Beyond mean-field approximation}
The lowest energy states obtained from the mean-field formulation in the previous section describe approximate ground states of the system. Specifically, we obtain two solutions, namely, $|\psi_N^{+}\rangle$ for $\{\alpha=\alpha_{+},\beta=\beta_{+}\}$ and  $|\psi_N^{-}\rangle$ for $\{\alpha=\alpha_{-},\beta=\beta_{-}\}$.  To improve ground state approximation, we consider a quantum superposition of these states, called Sch\"odinger cat states. It is more accurate  to describe ground state of the system.  The two possible ortho-normal ground states obtained by a quantum superposition of the two macroscopically distinct distributions $|\psi_N^+\rangle$ and $|\psi_N^-\rangle$ are 
given by \cite{Cirac}
\begin{equation}
|\psi^{\pm} \rangle = \frac{1}{\sqrt{2(1 \pm \Lambda^N)}}(|\psi_N^+ \rangle \pm |\psi_N^- \rangle).
\end{equation}
We treat $\{|\psi^- \rangle),|\psi^+ \rangle)\}$ as a basis of single qubit state such that 
\begin{equation}
|0\rangle \equiv |\psi^+\rangle,\,\,\,{\rm and}\,\,\,
		|1\rangle \equiv |\psi^-\rangle.
\end{equation}
The energies of the states obtained from 
$E_0=\langle \psi^+|\hat{H}_0|\psi^+\rangle$ and $E_1=\langle \psi^-|\hat{H}_0|\psi^-\rangle$ are given by
\begin{eqnarray}
E_0&=&\frac{1}{4(1+\Lambda^N)}\big[({}^N P_3) u_0(4-3\Lambda^2 +\Lambda^3)+4 ({}^N P_2)(u_0+u_1)
\nonumber\\&+&2u_2{}^N P_2\Lambda^2+4 N k_0p_0(1+\Lambda)\big]
\end{eqnarray}
and
\begin{eqnarray}
E_1&=&\frac{1}{4(1-\Lambda^N)}\big[({}^NP_3) u_0(4-3\Lambda^2 - \Lambda^3)
\nonumber\\&+&4 ({}^N P_2) (u_0 + u_1)(1-\Lambda^2)
\nonumber\\&+&4Nk_0 p_0(\Lambda -1)\big]
\end{eqnarray}
Figures 4 and 5 display the variations of ground state energy $E_0$ and the energy difference of the two low-lying  levels $(E_1-E_0)$ as functions of $k_0/N$, calculated using the beyond mean-field approximation. We see that the variation of $E_0$ shows similar trends in both numerical and beyond mean-field approaches.  Interestingly, the energies of $E_0$ and $E_1$ remain degenerate up to certain values of $k_0/N$. This is again similar to the result obtained from numerical simulation with a different range of $k_0/N$ for the degeneracy. Therefore, these two quasi-degenerate states can be used to encode qubits.
\begin{figure}[h!]
\centering
\includegraphics[width=8.0cm, height=5.5cm] {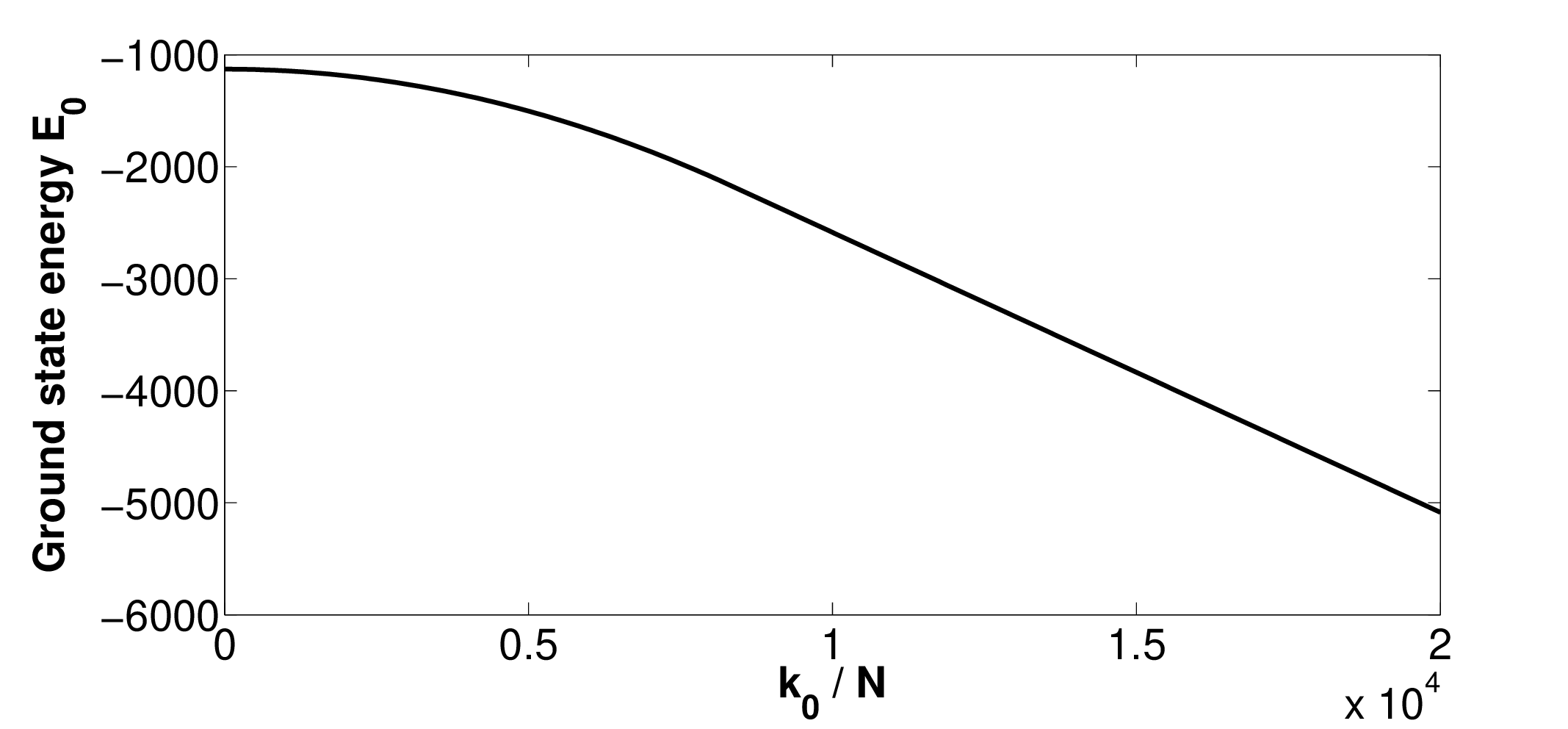}
\caption{Variation of ground state energy $E_0$ as a function of $k_{0}/N$ derived from the Schr\"odinger cat states.  The parameters taken are same as earlier: $u_0 = -3.35\times 10^{-6}, u_1= -2.85\times 10^{-3}, u_2 = 3.0\times 10^{-3}$ and $p_{0}= -1\times 10^{-6}$.}
\label{fig4}
\end{figure}
\begin{figure}[h!]
\centering
\includegraphics[width=8.0cm, height=5.5cm] {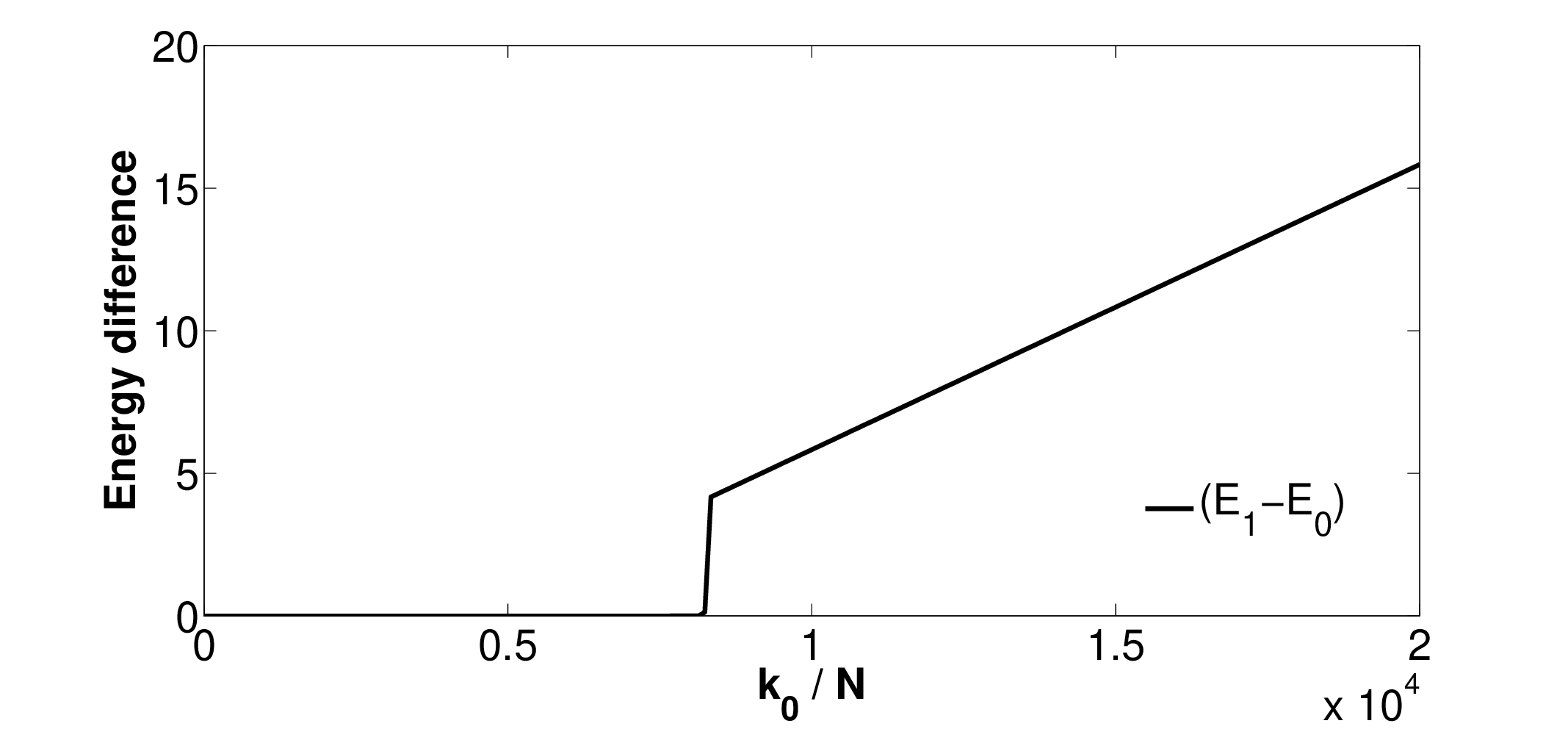}
\caption{Variation of the energy difference between two low-lying quasi degenerate energy levels as a function of $k_{0}/N$ derived from the Schr\"odinger cat states.  The parameters taken are same as earlier:  $u_0 = -3.35\times 10^{-6}, u_1= -2.85\times 10^{-3}, u_2 = 3.0\times 10^{-3}$ and $p_{0}= -1\times 10^{-6}$.}
\label{fig5}
\end{figure}
\begin{figure}[h!]
\centering
\includegraphics[width=8.2cm, height=5.2cm] {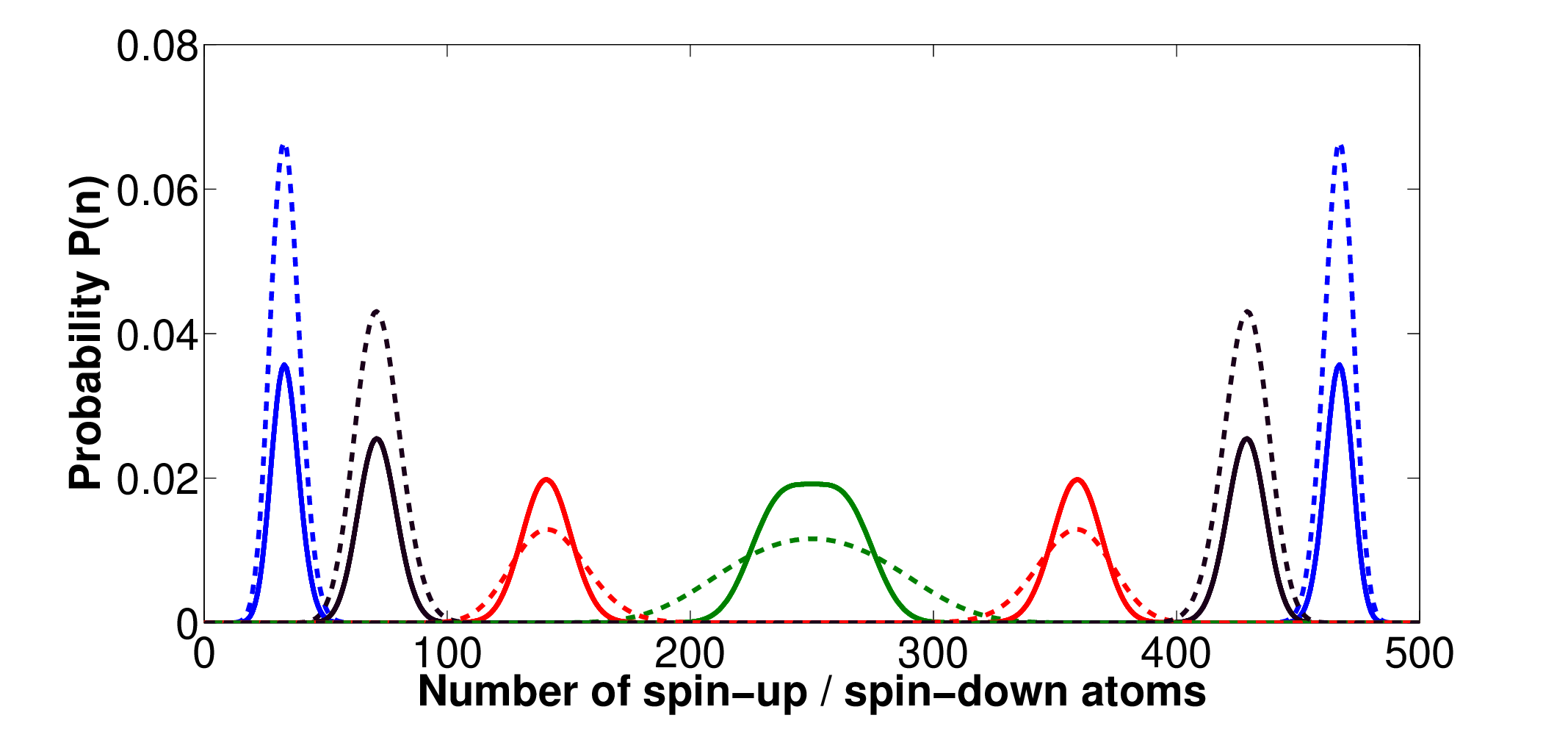}
\caption{Probability distribution for the ground state: exact versus beyond mean-field theory as a function of spin-up / spin-down atom number. Solid lines are the beyond mean-field calculations for $\Lambda=0.5$ (blue), $\Lambda=0.7$ (black), $\Lambda=0.9$ (red) and $\Lambda=0.998$ (green). Dashed lines are exact calculations for the corresponding $k_0$ values. Other parameters are same as in Fig.2.}
\label{fig6}
\end{figure}
From  the mean-field approximation we see that the probabilities of occupation in two spin states are equal as $\Lambda \,\rightarrow 1$. However, these probabilities become unequal as $\Lambda <1$. In Fig.6, we plot the results obtained from beyond mean-field approximation together with the exact numerical calculation on the variation of occupation probability with the number of spin-up / spin-down atoms for the same $\Lambda$-values as shown in Fig.3. We see that the beyond-field approximation describes the probability of occupation more accurately than the mean-field approximation. It shows that the occupation probability contains two peaks if  $\Lambda < 1$ and the peaks start to get merged as $\Lambda$ approaches to $1$. This is consistent with result obtained from mean-field approximation. Thus we see that the occupation probability of the two states can be varied with the parameter $\Lambda$. Particularly, $\Lambda$ can be considered as a parameter to control the amplitudes $c_1$ and $c_2$ of the qubit state:
\begin{equation}
		|\psi^{q} \rangle=c_1 |\psi^+\rangle e^{-iE_0t/\hbar}+c_2 |\psi^-\rangle e^{-iE_1t/\hbar}.
\end{equation}
Here $|c_1|^2+|c_2|^2=1$.
{The symmetric and antisymmetric combinations of the two macroscopically distinct states form a pair of nearly degenerate eigenstates. These combined states define an effective two-level system, which are well separated from higher energy states. We, therefore, identify these two states as the logical qubit basis states for encoding of quantum information.}

\section{Single-qubit gates by higher-order nonlinear perturbations}
Qubit is the building block for quantum computing and  the essential basis of quantum information processing is the coherent manipulation of qubits  \cite{Nielsen, Wie}. Mathematically, a qubit is defined as a coherent superposition of two orthonormal states with continuous phase and amplitude parameters. To encode a qubit, we need to find a physical system that acts as an effective two-level system. As we observed in the earlier section, the two lowest-energy states of SOC-BEC system are quasi-degenerate and well separated in energy from the rest of the spectrum under specific conditions. Therefore, the subspace spanned by these two states remains unaffected from perturbations as long as they are significantly smaller than this energy gap. { Higher-order nonlinear perturbations used for gate generation can be engineered through (i) state-dependent optical lattices which modify inter-component overlap\cite{Jaksch}, (ii) microwave dressing fields which induce correlated hopping \cite{Goldman} and (iii) Floquet modulation which generates effective multi-body terms\cite{Eckardt}. Explicit three-body interaction effects have been experimentally and theoretically studied in ultracold gases\cite{Daley, Johnson}, offering a pathway to the quintic nonlinearities. Such techniques allow controlled realization of operators of the form as required for nonlinear gate operations.}
 
The Bloch sphere offers an elegant geometric representation of qubit states. Let a perturbation $\hat{v}_{m,n}$ is applied to the system. It's projection on the subspace formed by $|\psi^0\rangle$ and $|\psi^1\rangle$ produces  rotations about 3-axes in the Bloch sphere. Particularly, the projection of $\hat{v}_{m,n}$ is given by
\begin{equation}
		\hat{P}\hat{v}_{m,n}\hat{P}=\Omega_0\boldsymbol{I}+\Omega_x\boldsymbol{\sigma_x}+\Omega_y\boldsymbol{\sigma_y}+\Omega_z\boldsymbol{\sigma_z}
\end{equation}
where  $\hat{P}=|\psi^0\rangle \langle \psi^0| + |\psi^1 \rangle \langle \psi^1 |$  being the projection operator and $\boldsymbol{I}, \boldsymbol{\sigma_x}, \boldsymbol{\sigma_y}, \boldsymbol{\sigma_z}$ are identity and Pauli matrices respectively while the coefficients $\Omega_x, \Omega_y$ and $\Omega_z$ give the rotations about $x, y$ and $z$ axes respectively. Note that the coefficient $\Omega_0$ increases the energy of the system only. The unitary operator corresponding to the perturbation is given by \cite{Preskill}
\begin{equation}
		\hat{U}(\Delta t)=\exp[-i(\hat{P}\hat{v}_{m,n}\hat{P})\Delta t/\hbar].
\end{equation}
Here $\Delta t$ represents duration of perturbation. This is the generalized form of single-qubit gates.

We know that two-and three-body scatterings are the dominant processes of interactions in ultracold atomic systems. This gives rise cubic and quintic nonlinearities. However, there are always some possibilities of higher-order interaction due to more than three-body scattering.  We treat the higher-order nonlinearity as perturbation in the following and find its effects on the qubit state.

First we take higher-order inter-component interaction as perturbation.  For such perturbation, $\hat{v}_{m,n}$  is given by
\begin{equation}
		\hat{v}_{m,n}=\lambda_0(\hat{a}^\dagger)^{m} \hat{a}^{m}(\hat{b}^\dagger)^{n} \hat{b}^{n}
\end{equation}
with $m, n \leq N$.  Here $\lambda_0$ is the strength of perturbation.  Understandably, the terms $(\hat{a}^\dagger)^m \hat{a}^m$ and $(\hat{b}^\dagger)^n \hat{b}^n$ represent effective $m$-body and $n$-body interactions among the atoms in a level and thus $(\hat{a}^\dagger)^m \hat{a}^m (\hat{b}^\dagger)^n \hat{b}^n$ gives generalized inter-component interaction. For $m=n=1$ it coincides with the cubic inter-component interaction.  The projection of this potential is given by
\begin{equation}
	\hat{P} \hat{v}_{m,n} \hat{P} = \Omega_0 \boldsymbol{I} + \Omega_x \boldsymbol{\sigma_{x}} + \Omega_z \boldsymbol{\sigma_{z}},
\end{equation}
where
\begin{eqnarray}
	\Omega_0(\Lambda)&=&\frac{\lambda_0 {}^N P_{(m+n)}}{2(1-\Lambda^{2N})}\big[\alpha_+^{2m}\beta_+^{2n}+\beta_+^{2m}\alpha_+^{2n}
	\nonumber\\&-&\frac{\Lambda^{2N}}{2^{m+n}}\big],
\end{eqnarray}
\begin{eqnarray}
	\Omega_x(\Lambda)&=&\dfrac{\lambda_0 {}^N P_{(m+n)}}{2\sqrt{1-\Lambda^{2N}}}\big[\alpha_+^{2m}\beta_+^{2n}-\beta_+^{2m}\alpha_+^{2n}\big],
\end{eqnarray}
\begin{equation}
	\Omega_y=0
\end{equation}
and
\begin{eqnarray}
	\Omega_z(\Lambda)&=&\frac{\lambda_0 {}^N P_{(m+n)}\Lambda^N}{2(1-\Lambda^{2N})}\big[\frac{1}{2^{m+n+1}}
	\nonumber\\&-&(\alpha_+^{2m}\beta_+^{2n}+\beta_+^{2m}\alpha_+^{2n})\big].
\end{eqnarray}
 Substituting Eqs.(19)-(23) in Eq.(17), we get the following unitary operator.
\begin{eqnarray}
		 \hat{U}(\Delta t,\Lambda)&=&\boldsymbol{I}\cos(\frac{\Omega_x \Delta t}{\hbar})\cos(\frac{\Omega_z\Delta t}{\hbar})
        \nonumber\\&-&i\boldsymbol{\sigma_x} \sin(\frac{\Omega_x\Delta t}{\hbar})\cos(\frac{\Omega_z\Delta t}{\hbar})
        \nonumber\\&+&i\boldsymbol{\sigma_y}\sin(\frac{\Omega_x\Delta t}{\hbar})\sin(\frac{\Omega_z\Delta t}{\hbar})
		\nonumber \\&-&i\boldsymbol{\sigma_z}\cos(\frac{\Omega_x\Delta t}{\hbar}) \sin(\frac{\Omega_z\Delta t}{\hbar}).
	\end{eqnarray}
In writing Eq.(24) we have omitted the global phase factor $\exp[-i\Omega_0 \Delta t/\hbar]$.
The unitary operator represents a composite rotation of a qubit. It is a single-qubit gate that performs rotations about multiple axes simultaneously. This  can act as a universal single-qubit gate. A specific rotation can be achieved by carefully controlling the coefficients $\Omega_x$ and $\Omega_z$ and taking an appropriate  duration ($\Delta t$) of the perturbation. For example, a $x$-rotation can be obtained by making the coefficient $\Omega_z=0$. Similarly, a phase gate can be generated by making the coefficient $\Omega_x=0$. {The relative phase between the qubit basis states is given by $\phi(t)=\frac{\Delta E }{\hbar}t$, where $\Delta E=|E_1-E_0|$. So, phase accumulation requires $\Delta E = E_{1} - E_{0} \neq 0$. Thus, to acquire a relative phase between $\Ket{0}$ and $\Ket{1}$ the degeneracy must be lifted. To initialize the qubit in the state $\Ket{0}$, the system must be prepared in its ground state, which lies in the quasi-degenerate regime $\Lambda \le 1$. Direct cooling in this regime is challenging due to the small energy gap between the two lowest states. Instead, the system is first cooled to near-zero temperature in the regime $\Lambda \ge 1$, where the energy gap is sufficiently large to ensure ground-state preparation. The parameter $\Lambda$ is then adiabatically reduced to reach the desired quasi-degenerate regime.}

In the second  case,  we take intra-component  interaction as perturbation. A generalized form of the perturbation is given by
\begin{equation}
\hat{v}_{m,n}=\lambda_0[(\hat{a}^\dagger)^{m} \hat{a}^{m}+(\hat{b}^\dagger)^{n} \hat{b}^{n}]
\end{equation}
with $m, n\leq N$. For $m=n=2$ and $m=n=3$, it gives cubic and quintic intra-component interactions. The projection of this perturbation can be calculated from Eq.(16) using 
\begin{eqnarray}
\Omega_0(\Lambda)&=&\frac{\lambda_0}{2(1-\Lambda^{2N})}\big[{}^N P_{m}(\alpha_+^{2m}+\beta_+^{2m}-\frac{\Lambda^{2N}}{2^{m}})
	\nonumber\\&+&{}^N P_{n}(\alpha_+^{2n}+\beta_+^{2n}-\frac{\Lambda^{2N}}{2^{n}})\big],
\end{eqnarray}
\begin{eqnarray}
\Omega_x(\Lambda)&=&\dfrac{\lambda_0}{2\sqrt{1-\Lambda^{2N}}}\big[{}^N P_{m}(\alpha_+^{2m}-\beta_+^{2m})
	\nonumber\\&+&{}^N P_{n}(\beta_+^{2n}-\alpha_+^{2n})\big],
\end{eqnarray}
\begin{equation}
	\Omega_y=0
\end{equation}
and 
\begin{eqnarray}
	\Omega_z(\Lambda)&=&\frac{\lambda_0}{2(1-\Lambda^{2N})}\big[{}^N P_{m}\big(\Lambda^{N}2^{1-m}
	\nonumber\\&-&\Lambda(\alpha_+^{2m}+\beta_+^{2m})\big)+{}^N P_{n}\big(\Lambda^{N}2^{1-n}
	\nonumber\\&-&\Lambda(\alpha_+^{2n}+\beta_+^{2n})\big)\big].
\end{eqnarray}
The unitary operator is given by
\begin{eqnarray}
		 \hat{U}(\Delta t,\Lambda)&= &\boldsymbol{I}\cos(\frac{\Omega_x \Delta t}{\hbar})\cos(\frac{\Omega_z\Delta t}{\hbar})
        \nonumber\\&-&i\boldsymbol{\sigma_x} \sin(\frac{\Omega_x\Delta t}{\hbar})\cos(\frac{\Omega_z\Delta t}{\hbar})
        \nonumber\\&+&i\boldsymbol{\sigma_y}\sin(\frac{\Omega_x\Delta t}{\hbar})\sin(\frac{\Omega_z\Delta t}{\hbar})
		\nonumber \\&-&i\boldsymbol{\sigma_z}\cos(\frac{\Omega_x\Delta t}{\hbar}) \sin(\frac{\Omega_z\Delta t}{\hbar}).
\end{eqnarray}
Here also we neglect the global phase factor, $\exp[-i\Omega_0 \Delta t/\hbar]$. The unitary operator represents a composite rotation of single-qubit. This single-qubit gate performs rotations about multiple axes simultaneously. For higher-order perturbation in intra-component interaction  with $m\neq n$,   $x$-rotation  and $z$-rotation occurs in the Bloch sphere. However,  we  get only $z$-rotation for $m=n$. In this case, $\Omega_x=0$. From Eqs.(26) and (29), we conclude that one  can generate similar gate using either higher-order inter-component or intra-component interactions as perturbations.

In the third case, we consider higher-order correlated hopping as perturbation. This type of perturbation can be created by applying an external alternating fields.  This field flips $m$ number of down-spin particles to up-spin and $n$ number of up-spin particles to down-spin periodically. In one complete cycle of the external field, the perturbation can be expressed as
\begin{equation}
		\hat{v}_{m,n}=\lambda_0[(\hat{a}^\dagger \hat{b})^{m}+(\hat{b}^\dagger \hat{a})^{n}]
\end{equation}
where $\lambda_0$ is the strength of perturbation field and $m\leq N$,$n\leq N$. 	Note that, for $m=n=1$, we get the last term of Eq.(6).
For this perturbation  the projection  can also be calculated from Eq.(16) using 
\begin{eqnarray}
	\Omega_0(\Lambda)&=&\frac{\lambda_0}{(1-\Lambda^{2N})}\big[{}^N P_{m}(\alpha_+\beta_+)^{m}
	\nonumber\\&+&{}^N P_{n}(\alpha_+\beta_+)^{n}\big],
	\end{eqnarray}
\begin{equation}
		\Omega_x=0,
\end{equation}
\begin{eqnarray}
	\Omega_y&=&\frac{i \lambda_0}{2\sqrt{1-{\Lambda}^{2N}}}\left[^NP_m \Lambda^{(N-m)}\left(\alpha_{-}^{2m}-\alpha_{+}^{2m}\right)\right.\nonumber\\
&+&\left. ^NP_m \Lambda^{(N-n)}\left(\alpha_{+}^{2n}-\alpha_{-}^{2n}\right)\right],
\end{eqnarray}
\begin{eqnarray}
		\Omega_z(\Lambda)&=&\frac{\lambda_0}{2(1-\Lambda^N)}\big\{[{}^N P_{m}(\alpha_+^{2m}+\beta_+^{2m})\Lambda^{N-m}
		\nonumber\\&+&{}^N P_{n}(\alpha_+^{2n}+\beta_+^{2n})\Lambda^{N-n}](1-\Lambda^{N})
		\nonumber\\&-&2\Lambda^{N}[{}^N P_{m}(\alpha_+\beta_+)^{m}+{}^N P_{n}(\alpha_+\beta_+)^{n}]\big\}
\end{eqnarray}
The unitary operator for this perturbation is given by
{\begin{eqnarray}
\hat{U}(\Delta t,\Lambda)&=&\boldsymbol{I}\cos(\frac{\Omega_y\Delta t}{\hbar})\cos(\frac{\Omega_z\Delta t}{\hbar})
            \nonumber\\&-&i\boldsymbol{\sigma_x}\sin(\frac{\Omega_y\Delta t}{\hbar})\sin(\frac{\Omega_z\Delta t}{\hbar})
            \nonumber\\&-&i\boldsymbol{\sigma_y} \sin(\frac{\Omega_y\Delta t}{\hbar})\cos(\frac{\Omega_z\Delta t}{\hbar})
\nonumber \\&-&i\boldsymbol{\sigma_z}\cos(\frac{\Omega_y\Delta t}{\hbar}) \sin(\frac{\Omega_z\Delta t}{\hbar}).
\end{eqnarray}}
We neglect the global phase factor as earlier. The higher-order correlated hopping perturbation generates only $z$ rotation  for $m=n$ in Bloch sphere. So a phase gate can be obtained using this type of perturbation.

\section{Concluding Remarks}
A system with two nearly equal energy levels is basic requirement for generating qubit. Utracold many body system in the form of Bose-Einstein condensates can serve as a two level system and thus provides a platform for simulating qubit. {Two hyperfine levels of the atoms can be connected by a Raman transition. For a qubit platform to be feasible, it must remain robust under realistic experimental imperfections. The uncontrolled interaction of the quantum systems with the environment may lead to a loss of coherence, a process known as decoherence. Sources of decoherence may be thermal fluctuations, trap losses due to inelastic collisions, scattering with background thermal atoms or spontaneous light scattering. Most suitable platform among the spin-orbit coupled BECs is $^{87}\mathrm{Rb}$ having long coherence time, well-controlled scattering lengths. The atomic species for which BEC was realized for the first time is $^{87}\mathrm{Rb}$\cite{Anderson} and the two states can be condensed simultaneously \cite{Myatt}. Using two counter-propagating Raman lasers that couple hyperfine states, synthetic SOC is generated.} We consider spin-orbit coupled two level BECs within the framework of second quantization formalism and show that two low-lying states of the system under certain condition remain degenerate.  More specifically,  we find low-energy states using direct numerical simulation and mean-field  approximation and show that there exits a upper limit of Raman coupling for degeneracy of the state. The occupation probabilities  of the states depend crucially on Raman coupling and  thus it can be used to control population imbalance between the two states.  We confirm through numerical simulation that two quasi-degenerate Schr\"odinger-cat states are suitable candidates for dealing with qubit.
 
We apply three different nonlinear perturbations, namely, higher-order intra-and inter-component interactions, and hopping to the system. These perturbations produce  rotations  of  the Schr\"odinger-cat states on Bloch sphere. Rotation in the Bloch sphere is equivalent to a gate operation. Therefore, using different types of perturbation, we can realize various quantum gates in spin-orbit coupled Bose-Einstein condensates. It is shown that the Raman coupling plays a significant role in controlling  quantum gate through changing occupation probability of the spin states. Quantum gates based on rotation about any axis can be generated by changing the Raman coupling between atomic states adiabatically and adding a state-dependent perturbative potential which causes  energy difference between the atomic states.

{For a system prepared in the initial state $\Ket{00..0}$, successful quantum gate operations require that the dynamics of the system  (i.e. the dynamics due to the gates) occur on timescales much shorter than the decoherence time of the system. This is expressed by the condition $t_g \ll \tau$, where $t_g$ is the typical timescale of a quantum gate and $\tau$ is the decoherence time of the physical system (e.g. for a gate generating a rotation of frequency $\Omega$ by an angle $\varphi$, the required time is given by $t = \varphi / \Omega$). For a typical experimental scales Raman coupling $\Omega_{R} \sim 1 - 10\,\mathrm{kHz}$ \cite{Lin, Zhang2012}, interaction energy $\sim 10 - 100\,\mathrm{Hz}$\cite{Pitaevskii} and coherence time $\tau \sim 100$ ms\cite{Stamper}, the resulting single-qubit gate times $t_{g} \sim 1 - 10$ ms, well below decoherence limits, making nonlinear gates experimentally feasible.}

{The above analysis demonstrates that higher-order nonlinearities in spin-orbit coupled Bose-Einstein condensates can help stabilize qubit subspaces and enable controllable gate operations. The present cutting-edge ultracold atom experiments can achieve the parameter regimes needed for quintic-assisted qubit manipulation via Raman-induced SOC methods, Feshbach-tuned interactions, and lattice-enhanced multi-body effects. Therefore, higher-order nonlinear interaction improves the robustness of the effective qubit manifold in addition to being helpful for generating gates.}
\subsection*{ACKNOWLEDGEMENT}
 Prithwish Ghosh would like to thank ‘West Bengal Higher Education Department’ for providing Swami
Vivekananda Merit-Cum-Means Scholarship with F. No. WBP241723462687.

\end{document}